# Microstructure and high critical current of powder in tube $MgB_2$

A. Serquis[a], L. Civale, D. L. Hammon[b], J. Y. Coulter, X. Z. Liao, Y. T. Zhu, D. E. Peterson, and F. M. Mueller

Superconductivity Technology Center, MS K763, Los Alamos National Laboratory, Los Alamos, NM 87545, USA

We report dc transport and magnetization measurements of $J_c$ in $MgB_2$ wires fabricated by the powder-in-tube method, using commercial $MgB_2$ powder with 5 %at Mg powder added as an additional source of magnesium, and stainless steel as sheath material. By appropriate heat treatments, we have been able to increase $J_c$ by more than one order of magnitude from that of the as-drawn wire. We show that one beneficial effect of the annealing is the elimination of most of the micro-cracks, and we correlate the increase in $J_c$ with the disappearance of the weak-link-type behavior.

---

[a] Electronic mail: aserquis@lanl.gov

[b] Materials Technology Metallurgy Group, MS G755, Los Alamos National Laboratory

The great interest in superconducting $MgB_2$ was initially driven by its high transition temperature $T_c$ (~ 39 K), chemical simplicity and low cost[1]. Expectations were further raised by the realization that randomly oriented polycrystalline samples may be free from weak-link behavior at the grain boundaries (GB). However, the ceramic-like properties of $MgB_2$ complicate the production of wires or tapes required for practical applications. Indeed, although considerable progress has already been made, it is still true that the best critical current densities $J_c$ reported so far on $MgB_2$ wires or tapes fabricated using the powder-in-tube (PIT) method are well below those of bulk samples prepared under high pressure.[2,3,4]

To optimize $J_c$ in $MgB_2$ wires and tapes, we need a deeper understanding on how fabrication and processing conditions affect microstructure and vortex pinning. Some general facts have been established. For instance, the use of sheath materials with high toughness and chemical compatibility such as Fe or stainless steel[5,6,7] is essential to achieve high $J_c$ while Cu, Ni or Ag sheaths lead to lower values.[6,8,9] Also, some post-annealing appears to be necessary to achieve higher $J_c$.[5,7,10] Finally, tapes show higher $J_c$'s than round wires, due mainly to reduced porosity.[5,6,11,12,13] However, except for these basic issues little else is known about current limiting mechanisms in $MgB_2$.

The first fundamental question is which structural features are the relevant pinning centers. It has been suggested[14] that GB may be the main source of vortex pinning in $MgB_2$. Although this is plausible, the experimental evidence is inconclusive. If GB are indeed effective pins, to optimize $J_c$ one should make the grains as small and randomly oriented as possible. This is the opposite to the case of the high temperature superconductors, where GB are weak links whose detrimental effects are reduced by making grains as large and aligned as possible. Another source of pinning are intra-grain defects. We have recently shown[15,16] that bulk $MgB_2$ prepared by solid-state reaction contains a large amount of intra-grain $Mg(B,O)_2$ precipitates coherent with the $MgB_2$ matrix, with sizes ranging from 5 nm to 100 nm, which are very well suited to act as pinning centers. A second and equally important requirement for wire and tape optimization is a better understanding of more macroscopic aspects such as the influence of the porosity, the formation of MgO phase, the presence of microcracks, and the weak link behavior after deformation.

In this letter we report on the fabrication of $MgB_2$ wires by PIT using stainless steel as sheath material. $J_c$ was determined from dc transport and magnetization. At 4 K, un-annealed wires exhibit a $J_c \sim 50000$ A/cm$^2$ at zero field and 1000 A/cm$^2$ at 5 T. By appropriate heat treatments, we have been able to increase $J_c$ by more than one order of magnitude. Microstructural analyses show that the most important beneficial effect of the annealing is the elimination of the micro-cracks present in the un-annealed wire. We also discuss the influence of grain size and porosity. Although the largest $J_c$'s in $MgB_2$ are likely to be obtained in tapes, their optimization will require the exploration of a large additional set of post-drawing processing parameters. Thus, as our present goal is to investigate the correlation among processing, microstructure and pinning, we confined this study to round wires.

Commercial $MgB_2$ powder was ball-milled for two hours and packed into stainless steel tubes (inner and outer diameters were 4.6 and 6.4 mm, respectively) in an Ar atmosphere, adding 5% at Mg powder as an extra source of magnesium. The tubes were cold-drawn into round wires with a final external diameter of 1.4 mm, with an intermediate annealing. The surface morphology and microstructures of the samples were investigated using a JEOL 6300FX scanning electron microscope (SEM) and a Philips CM30 transmission electron microscope (TEM) operated at 300 kV. SEM observations and energy dispersive spectrum (EDS) analysis showed no reaction between the $MgB_2$ and the stainless steel (see inset of Fig. 1).

We measured the dc transport critical current $I_c$ at 4 K, with the wires immersed in liquid He. We used samples ~10 cm long with voltage contacts placed 2 to 3 cm apart, in order to eliminate the initial ohmic behavior that we sometimes observed in the I-V curves of shorter wires. That artifact is due to poor connectivity between the $MgB_2$ and the stainless steel, which results in a long current transfer. We defined $I_c$ using a 1 μV criterion. A Quantum Design SQUID magnetometer was used to measure the dc magnetization M of ~ 4-5 mm long pieces of the wires, with H parallel to the wire axis.

Fig. 1 shows $I_c$ as a function of applied magnetic field H for one piece of our starting several-meters-long un-annealed wire, identified as A-0. Both the values of $J_c$ (shown on the left-side scale of the figure) and the H dependence are quite acceptable for an as-drawn $MgB_2$ wire. Another non-adjacent piece of A-0 exhibits a very similar Ic(H). This indicates that a long length of good quality and homogeneous wire was available for our annealing studies. Also shown in Fig. 1 are the data for two annealed wires, A-1 and A-2, that will be discussed later.

Fig. 2 shows $4\pi M/H$ vs. temperature at H = 20 Oe for pieces of the three wires shown in Fig. 1. Both the zero field cooling (ZFC) and field cooling (FC) curves are shown. The $T_c \sim 37.5$ K is the same in all cases and only slightly lower than in bulk samples. The ZFC curve for A-0 exhibits the two-steps typical of weak-link behavior. At the lowest temperatures $4\pi M/H \sim 1$, indicating that the screening currents flow through the whole sample, but connectivity is gradually lost as T increases and the weak links appear to collapse at T ~ 29 K (see arrow). Although the substantial $J_c$ values of A-0 at high fields (see Fig. 1) indicate that some fraction of the current paths does not contain weak links, it is also clear from Fig. 2 that the first goal of further processing should be to improve connectivity. To that end, we have performed a study of the effects of annealing conditions on microstructure and $J_c$. Here we will summarize the most relevant findings, more detailed results will be presented elsewhere. [17]

The data for wire A-1, shown in Figs. 1 and 2, is an example of an inappropriate heat treatment. In this case a 10 cm long wire was annealed in high purity flowing Ar. The temperature was increased at a rate of 10°C/min, then maintained at 900°C for 15 minutes, slowly cooled to 500°C at a rate of 0.5°C/min, then furnace cooled down to room temperature. While Fig. 1 shows that the transport $J_c$ has decreased by an order of magnitude, the reduced screening fraction and the absence of the two steps in the ZFC curve in Fig. 2 indicate that the current carrying capability of the weak links has deteriorated.

A sharply contrasting result was obtained for A-2 wire. Here a 6 m long wire was annealed in vacuum. Temperature was increased at a rate of 35°C/min, kept at 900°C for 30 minutes, and gas quenched with Ar. Fig. 1 shows the transport $J_c$ for a piece cut from A-2 after annealing, which is 20 times higher than the pre-annealing values. Only data above 4.75 T are shown, because at lower fields, $I_c$ was so high that strong heating at the current contacts precluded the collection of reliable data. The dotted line in Fig. 1 is an extrapolation assumimg the same H dependence as in A-0. Four other pieces of A-2 were also measured and showed similar results (within 20%). Consistently, the ZFC data in Fig. 2 exhibit almost perfect screening with no hint of weak-link behavior, confirming that the $J_c$ increase is mainly due to the improved connectivity. Also shown in Fig. 1 is the $J_c$ of A-2, as calculated from magnetization loops M(H) using the critical state Bean model and assuming a uniform $J_c$ flowing through the entire sample. The similar H dependence of transport and magnetization $J_c$ (except near H=0, where self-field effects dominate), and the coincidence within a factor of 2 in the field region of overlap, demonstrate that in the transport measurements the current is flowing through most of the wire cross section.

Typical microstructures observed by TEM are shown in Fig. 3, where (a) and (b) are from A-0, (c) and (d) from A-1, and (e) and (f) from A-2. The distribution of grain sizes can be estimated from dark field images such as those in Figs. 3 (a), (c) and (e). The observations can be summarized as follows: (i) in all three samples, most $MgB_2$ crystals are of sub-micrometer sizes, and in some rare cases, the grain size can reach around one micron [e.g., the largest bright area in Fig. 3(c)]; (ii) a large density of grains even smaller than 0.1 μm is seen in A-0 and A-1, while A-2 has fewer of those; (iii) while A-0 and A-1 have similar grain size distribution, A-2 has a slightly larger average grain size. Several conclusions can be extracted: Firstly, if pinning by GB is important, small grain size is desirable; Secondly, the lower $J_c$ of A-1 with respect to A-0 is not due to a reduction of GB pinning, as the grain size distribution is the same in both cases; Thirdly, the $J_c$ increase in A-2 is not due to an increase of GB pinning. If anything, GB pinning in A-2 must be smaller than in A-0, because the average grain size is larger. This last fact should by no means be taken as evidence against GB pinning. The grain size in A-2 is still small and is thus appropriate for GB pinning, and in any case the grain size differences are too small to draw a clear conclusion. As already mentioned, the main issue here is connectivity.

Figs. 3 (b) and (d) show examples of micro-cracks in A-0 and A-1 respectively. These micro-cracks extend over distances much larger than the grain size, and thus introduce a strong limitation to the flow of currents. Micro-cracks were observed both in A-0 and A-1, but *they appear to be absent in* A-2. From this we conclude that *the main beneficial effect of the annealing of* A-2 *was the elimination of the micro-cracks*. Finally, Fig. 3 (f) shows pores in A-2. A substantial and similar density of pores is seen in the three samples, indicating that the reduction of porosity was not a relevant factor for the $J_c$ improvement in A-2.

To determine the reason for the bad connectivity in A-1, we annealed a ~ 5 cm piece of A-0, under similar conditions as in A-1 but for a shorter time. We then cut a series of ~ 5 mm long consecutive samples and measured their magnetization. We found[17] that both the connectivity (as parameterized by the double step ZFC as that of A-0 in Fig. 2) and the $J_c$ were worst for the end samples and systematically improved in samples taken closer to the center. This indicates that the deterioration was due to a loss of Mg through the wire ends during annealing. The loss of Mg was precluded in A-2 for two reasons. First, the wire was very long, so only a small length at each end lost Mg. Second, the heating was very fast, thus reducing the Mg loss before the recrystallization temperature was reached.

To improve the connectivity, annealing temperatures above 800-850°C are necessary. [14] On the other hand, heat treatments increase the grain size, so if pinning by GB is relevant the annealing times should not be too long or annealing temperatures not too high.[9,10] Processing conditions required to increase the density of $Mg(B,O)_2$ precipitates must also be explored. Finally, the pores still visible in A-2 are detrimental for $J_c$. All these considerations indicate that the optimization of annealing and processing conditions may be challenging due to conflicting constrains.

In summary, the $J_c$ in our wires was increased by a factor of 20 by appropriate annealing. The improvement was mainly due to the elimination of micro-cracks. To our knowledge, our $J_c$ is the highest reported in a round wire, and only slightly lower than the best result in tapes.[5] In addition, our wires are very sensitive to the heat treatments ($J_c$ in A-1 and A-2 differs by two orders of magnitude), probably due to the Mg excess in the initial powder. It is thus quite possible that we are still far away from the optimum processing conditions and a significant pinning improvement can still be achieved.

Figure captions

Figure 1- $J_c$ vs. applied field at 4 K as measured by transport for wires A-0, A-1 and A-2 (full symbols) and by magnetization for wire A-2 (open symbols). Extrapolated $J_c$ for A-2 is indicated by the dotted line. Inset: typical cross section of the wires as observed by SEM.

Figure 2 – Zero Field Cooling (ZFC) dc magnetization as a function of temperature for as-drawn and annealed $MgB_2$ wires. The arrow indicates the collapse of the weak links.

Figure 3 - TEM dark field [(a),(c),(e)] and bright field [(b),(d),(f)] images, where (a) and (b) are from A-0, (c) and (d) from A-1, and (e) and (f) from A-2.

# Figure 1

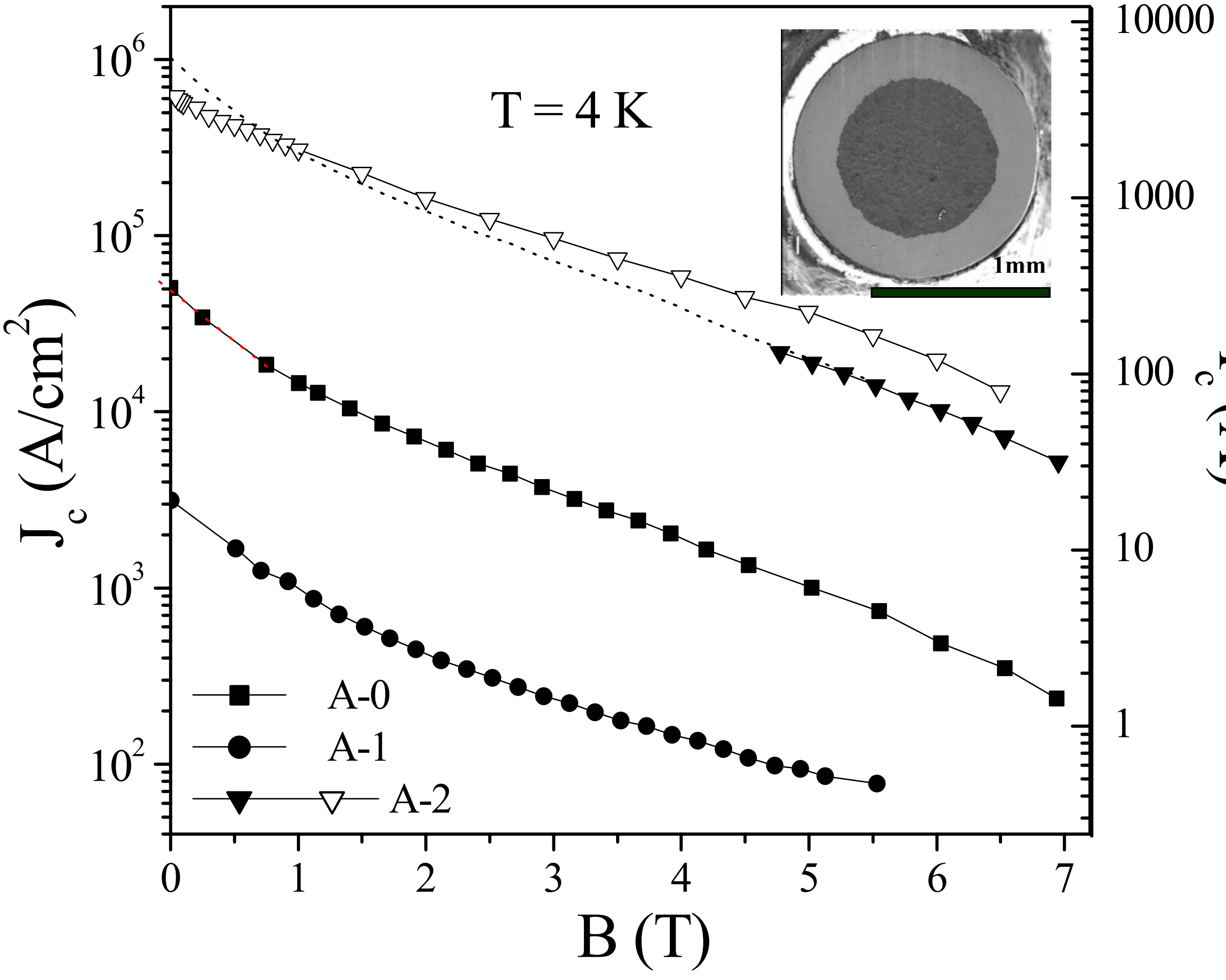

# Figure 2

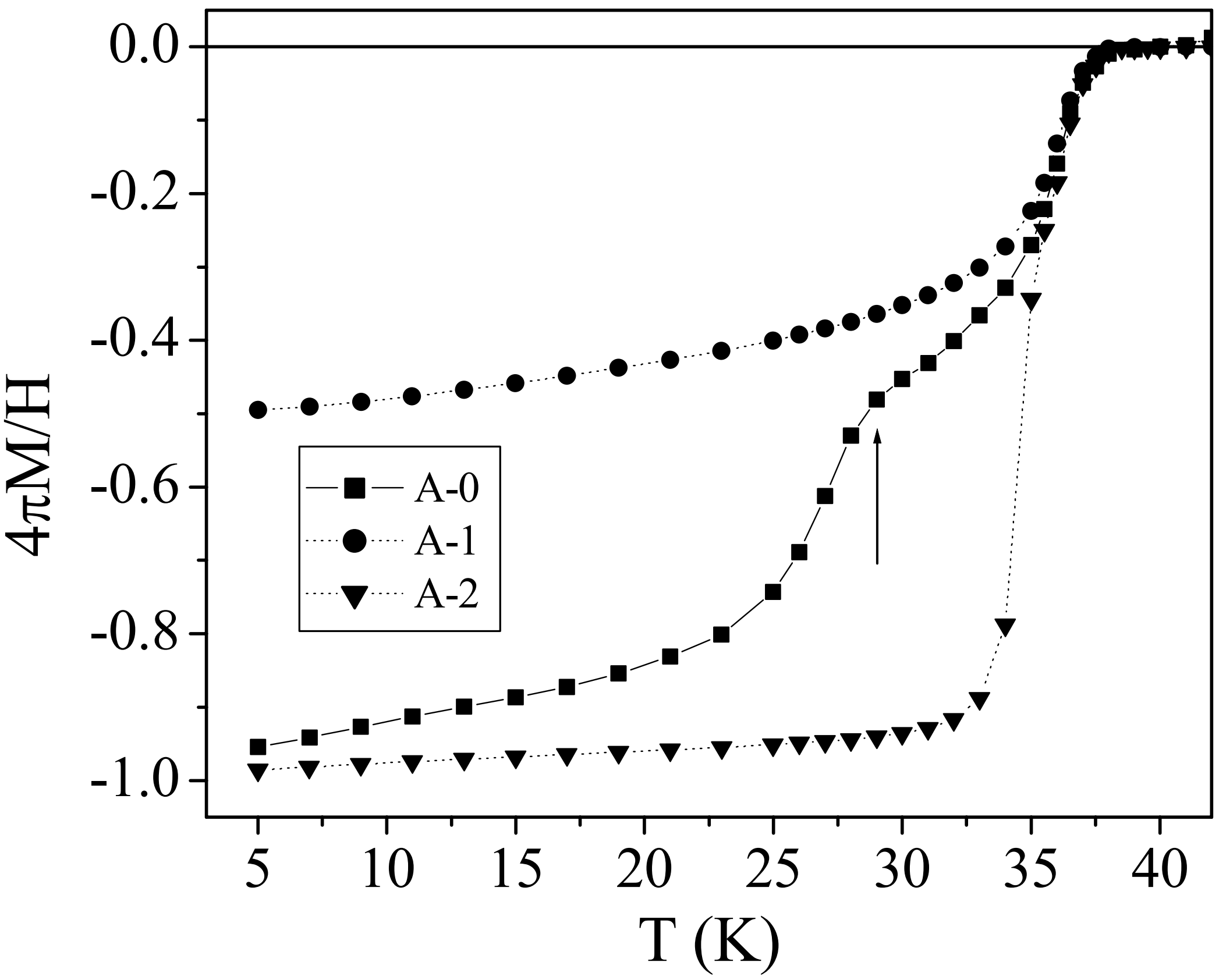

# Figure 3

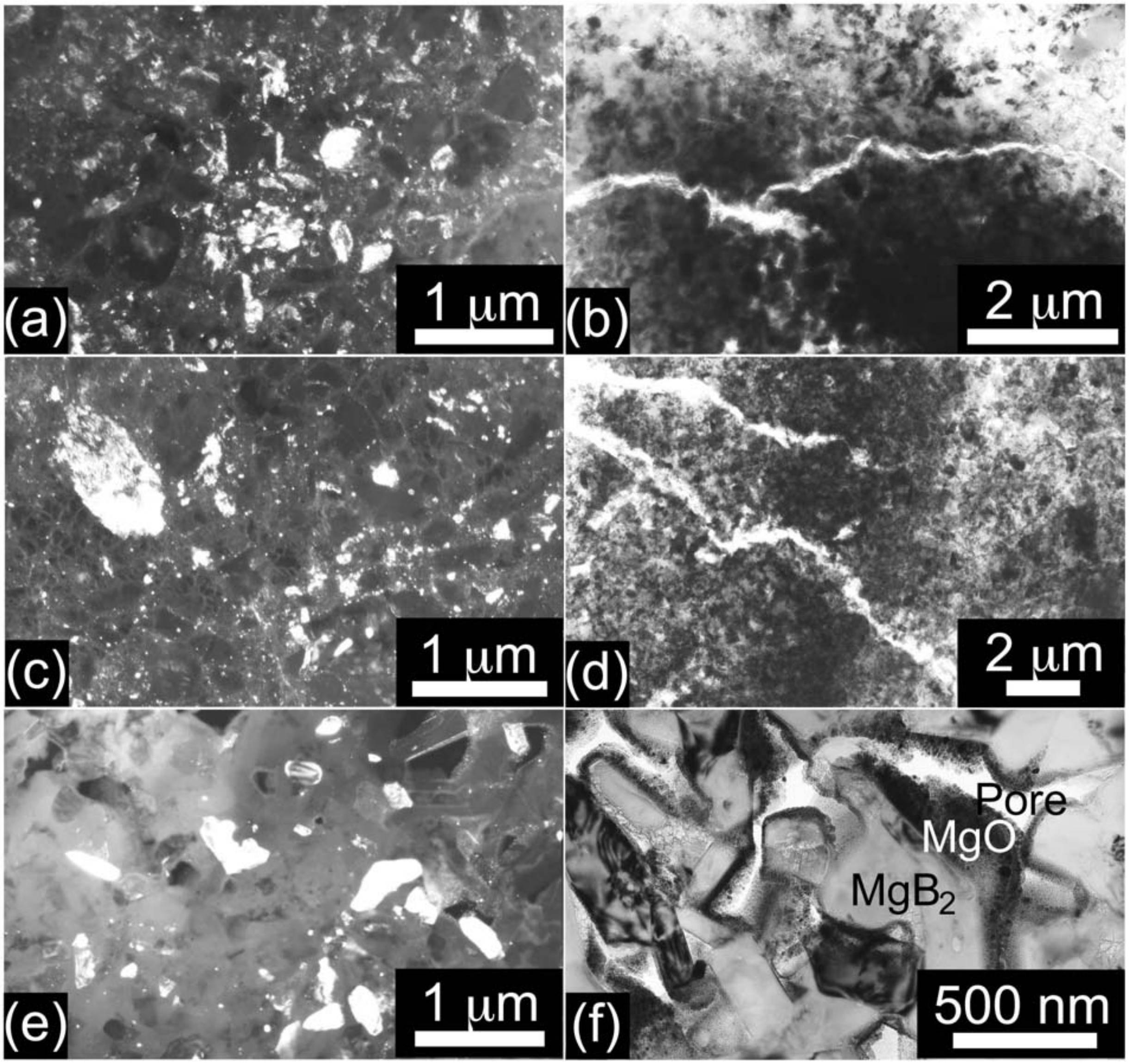